\documentclass{ws-ijmpe}
\usepackage[super,compress]{cite}
\usepackage[dvips,breaklinks]{hyperref}
\hypersetup{colorlinks,urlcolor=black,citecolor=black,linkcolor=black,filecolor=black}
\usepackage{breakurl}
\def\bib{B\kern-.05em{I}\kern-.025em{B}\kern-.08em}
\def\btex{B\kern-.05em{I}\kern-.025em{B}\kern-.08em\TeX}

\newcommand{\lwk}{{{\rm low}\mbox{-}k}}
\newcommand{\vlwk}{$V_{{\rm low}\mbox{-}k}$}
\newcommand{\thetaeff}{$\Theta_{\rm eff}$}

\newcommand{\zbb}{$0\nu\beta\beta$}
\newcommand{\dbb}{$2\nu\beta\beta$}
\newcommand{\heff}{$H_{\rm eff}$}
\newcommand{\heffs}{$H_{\rm eff}$'s}
\newcommand{\qbox}{$\hat{Q}$~box}
\newcommand{\tbox}{$\hat{\Theta}$~box}

\newcommand{\nmed}{$M^{2\nu}$}
\newcommand{\nmeds}{$M^{2\nu}$'s}

\begin{document}

\markboth{Authors' Names}{Instructions for typing manuscripts (paper's title)}

\catchline{}{}{}{}{}

\title{A glimpse into an effective world}

\author{Luigi Coraggio and Nunzio Itaco}

\address{Dipartimento di Matematica e Fisica, \\ 
Universit\`a degli Studi della Campania ``Luigi Vanvitelli'', \\ 
viale Abramo Lincoln 5, I-81100 Caserta, Italy, \\
and Istituto Nazionale di Fisica Nucleare, \\
Complesso Universitario di Monte  S. Angelo, \\
Via Cintia - I-80126 Napoli, Italy
\\
luigi.coraggio@na.infn.it\\
nunzio.itaco@na.infn.it}

\maketitle

\begin{history}
\end{history}

\begin{abstract}
Our contribution aims to celebrate the immeasurable contribution that
Tom Kuo has provided to the understanding of the structure of atomic
nuclei, and also of the infinite nuclear matter, in terms of the
fundamental principles governing the realistic nuclear potential. 
The authors want to testify Tom Kuo's heritage and impact on their
approach to the study of nuclear systems by reviewing some recent
findings on the role of the two-body component of shell-model
effective $\beta$-decay operators.
The focus is spotted on the so-called Pauli-blocking effect, that
plays a non-negligible role in nuclei characterized by a large number
of valence nucleons.
\end{abstract}

\keywords{Nuclear shell model; effective interactions; $\beta$ decay.}

\ccode{PACS numbers:21.60.Cs, 21.30.Fe, 27.40.+z}


\section*{Dedication}
I have collaborated with Tom for about twenty years, since the start
of PhD research work, and there is not a single page of the published
papers I have signed that have been not strongly influenced by his
teachings and his approach to the study of nuclear structure.

I met him in person, for the first time, in the spring of 1995, and,
among many topics we discussed in those days, a sentence of his has
been impressed in my mind: {\it everything we see is "effective"!}

I would like to think that this was his basic approach to physics,
namely that our reality is an ``effective'' representation of
underlying theories that we cannot directly manage to interpret the
observables, a viewpoint that reconnects to Plato's Allegory of the
Cave.

As a matter of fact, most of Tom's research has been devoted to develop
a theoretical framework where realistic nuclear potentials could be
employed to study the structure of nuclear systems, providing
many-body methods that could be able, at the same time, to reproduce
observables as well as to be valuable predictive tools.
He had the clear vision that, because of the intrinsic nature of the
nuclear potential that drives to highly-correlated systems, the most
efficient and profitable way to tackle the many-body problem should be
the reduction of the degrees of freedom of the problem, retaining only
those that would be crucial to describe the most relevant features of
nuclear systems.

Such a procedure had to be grounded on the theoretical construction of
new effective operators that would bridge the underlying theory -- the
nuclear potentials -- to the real world of the observables.

Tom's perspective on the nuclear many-body problem has been fully
successful, as can be testified by the results that his research
work has reached during a time period that spans from the second half of
1960s to the middle of 2010s, covering a large variety of scientific
interests and topics, always driven by an active curiosity to
understand the physics world with plain and clear answers to the
several questions the nature poses.

{\it L. Coraggio}

\section{Introduction}\label{intro}
Throughout his scientific career, Tom Kuo has extensively
studied the problem of the derivation of the shell-model effective
Hamiltonian from realistic nuclear potentials, in terms of the
many-body perturbation theory \cite{Kuo90,Hjorth95,Coraggio09a}.

His contributions to such a topic started from the seminal paper of the
Kuo-Brown residual interaction for $sd$ nuclei \cite{Kuo66}, and
subsequently focused to develop the folded-diagram method to build
shell-model effective Hamiltonians by way of the perturbative
expansion of the \qbox~ vertex function \cite{Kuo71,Krenciglowa74}.

The success of his research work is testified by wide amount of
nuclear structure studies which have benefitted from shell-model (SM)
calculations that have been performed by employing residual two-body
potentials derived from realistic nucleon-nucleon ($NN$) potentials by
him and his coworkers.

In this paper, we consider another important aspect of SM
calculations that are carried out starting from realistic nuclear
potentials, namely, the derivation of effective decay and transition
operators which account for the degrees of freedom that are not
explicitly included by the truncated model space of the valence
nucleons.
Tom Kuo addressed this topic both to evaluate effective charges for
electric-quadrupole $E2$ transitions \cite{Krenciglowa77}, and
effective two-body decay operators for the calculation of neutrinoless
double-$\beta$ decay (\zbb) nuclear matrix elements \cite{Song91}.
Moreover, in a topical paper about the evaluation of Goldstone
diagrams in an angular-momentum coupled representation, he and his
coworkers developed the formalism to calculate the diagrams emerging
from a perturbative expansion of effective decay operators
\cite{Kuo81}.

Here, we will focus our attention on a particular aspect of the
perturbative expansion of SM effective decay operators
\thetaeff, namely on the contribution of many-body terms that should
be included when deriving effective operators of one-body processes
for nuclear systems with more than one valence nucleon.
These contributions are responsible of the so-called ``blocking
effect'', which accounts for the Pauli exclusion principle when the
number of interacting nucleons evolves filling the orbitals of the
model space.
As a matter of fact, the violation of the exclusion principle by using
$n$-body effective operators in a ($n+1$)-particle system is corrected
by ($n+1$)-body diagrams.

In general, it is well known that induced many-body components of
effective operators soften the action of the one-body one \cite{Ellis77};
for example the electric-quadrupole effective charges are reduced in
many-valence nucleon systems, as well as the quenching factors for
Gamow-Teller (GT) transitions are enlarged.

In present work, we have chosen as subject of our study, the
calculation of the two-neutrino matrix element \nmed~ of the
two-neutrino double-$\beta$ (\dbb) decay of $^{100}$Mo, that has been
already studied in the framework of the realistic shell model starting
from the high-precision $NN$ potential CDBonn \cite{Machleidt01b},
whose repulsive high-momentum components have been integrated out by
way of the \vlwk~ renormalization procedure \cite{Bogner02}.
Then, a SM effective Hamiltonian \heff~ has been derived to calculate the
nuclear wave functions of parent and granddaughter nuclei $^{100}$Mo,
$^{100}$Ru, as well as the one-body effective GT decay operator to
obtain the matrix element \nmed~\cite{Coraggio22a}.

For such a calculation the reference core that has been considered
is $^{78}$Ni, consequently the nuclei involved in the decay under
investigation are described in terms of 22 valence nucleons
interacting in a model space that is spanned by four proton orbitals
$0f_{5/2},1p_{3/2},1p_{1/2},0g_{9/2}$ and five neutron orbitals
$0g_{7/2},1d_{5/2},1d_{3/2},2s_{1/2},0h_{11/2}$.

This means that the SM effective Hamiltonian and decay operators have
to consider also induced many-body components that account for the
filling of the single-particle orbitals (blocking effect).
In terms of a perturbative approach to the derivation of \heff~ and
\thetaeff, at least three-body diagrams and two-body diagrams have to
be included in the perturbative expansion for two-body Hamiltonian and
single-particle operators, respectively, as the leading-order
contributions of the induced many-body components of the effective
operators.
As regards the role of induced three-body contributions to \heff, we
have investigated their impact on the prediction of the neutron drip
line of calcium isotopes \cite{Coraggio20e}, and included their
contribution in the \heffs~ which have been employed in several
shell-model studies
\cite{Ma19,Coraggio21,Coraggio22a,Coraggio24a,Coraggio24b}.

It has to be pointed out that, since currently it is not
computationally possible to consider three-body components of \heff~
in large model spaces and for many-valence nucleon systems, the way to
manage three-body terms of \heff~ is to normal order these
contributions with respect to the nucleus under investigation, as a
reference state \cite{HjorthJensen17}, and neglect the residual three-body
term of the normal ordering expansion.
This leads to derive a density-dependent two-body term from the
three-body contribution, and the details of such an approach can be
found in Ref. \cite{Coraggio20c}.

The paper is organized as follows.

In the following section, we report some of the details of our theoretical
approach to the description of double-$\beta$ decay of $^{100}$Mo,
namely the derivation of the SM effective Hamiltonian and
decay operators, spotting a particular attention on the two-body
contribution of \thetaeff.
Then, in Sec. \ref{results} we present the results we have obtained
for the calculated  GT-strength distributions and for the \nmeds~ of
$^{100}$Mo decay, and compare them with those we have obtained in our
previous study in Ref. \cite{Coraggio22a} as well as with the available
data.
Our conclusions are drawn in the last section.

\section{Theoretical framework}\label{outline}
\subsection{The SM effective Hamiltonian}\label{heff}
Our calculation starts from the high-precision CD-Bonn $NN$ potential
\cite{Machleidt01b}, whose repulsive high-momentum components are
renormalized by way of the \vlwk~ approach \cite{Bogner02,Coraggio09a}.
As is well known, the \vlwk~ unitary transformation provides a smooth
potential and at the same time reproduces all the values of the
two-body observables which could be obtained with the CD-Bonn
potential.

The \vlwk~ matrix elements are inserted as interaction vertices of the
perturbative expansion of \heff~ and \thetaeff, and most of the details
of such an approach can be found in
Refs. \cite{Hjorth95,Coraggio12a,Coraggio20c}.
In this section, we outline briefly our procedure to derive SM
effective operators.

As regards \heff, we start from the full nuclear Hamiltonian $H$ that,
within the nuclear shell model, is separated into a one-body component
$H_0$ plus the residual interaction $H_1$ by introducing an auxiliary
potential $U$:

\begin{eqnarray}
 H &= & T + V_\lwk = (T+U)+(V_\lwk-U)= \nonumber \\
~& = &H_{0}+H_{1}~.\label{smham}
\end{eqnarray}

The one-body Hamiltonian $H_0$ set up the SM basis, that in our case
is the harmonic-oscillator (HO) one, but the Hamiltonian $H$ cannot be
diagonalized for a $A$-nucleon system in an infinite basis of
eigenvectors of $H_0$.
Then, an effective Hamiltonian \heff~ is derived, which is active only
in a truncated model space that, in order to describe a system as
$^{100}$Mo, is spanned by four proton -- $0f_{5/2}, 1p_{3/2},
1p_{1/2}, 0g_{9/2}$ -- and five neutron orbitals -- $0g_{7/2},
1d_{5/2}, 1d_{3/2}, 2s_{1/2}, 0h_{11/2}$ -- outside $^{78}$Ni core.

To this end, we resort to the time-dependent perturbation theory by
way of the Kuo-Lee-Ratcliff folded-diagram expansion of \heff~ in
terms of the \qbox~vertex function \cite{Kuo71,Kuo90}:
\begin{equation}
H^{\rm eff}_1 (\omega) = \hat{Q}(\epsilon_0) - P H_1 Q \frac{1}{\epsilon_0
  - Q H Q} \omega H^{\rm eff}_1 (\omega) ~, \label{eqfinal}
\end{equation}
\noindent
where $\omega$ is the wave operator decoupling the model space $P$ and
its complement $Q=1-P$, and $\epsilon_0$ is the eigenvalue of the
unperturbed degenerate HO Hamiltonian $H_0$.

The \qbox~is defined as
\begin{equation}
\hat{Q} (\epsilon) = P H_1 P + P H_1 Q \frac{1}{\epsilon - Q H Q} Q
H_1 P ~, \label{qbox}
\end{equation}
\noindent
and $\epsilon$ is an energy parameter called ``starting energy''.

As for the diagonalization of the full Hamiltonian $H$, an exact
calculation of the \qbox~ is computationally prohibitive, then the term
$1/(\epsilon - Q H Q)$ may be expanded as a power series

\begin{equation}
\frac{1}{\epsilon - Q H Q} = \sum_{n=0}^{\infty} \frac{1}{\epsilon -Q
  H_0 Q} \left( \frac{Q H_1 Q}{\epsilon -Q H_0 Q} \right)^{n} ~.
\end{equation}

\noindent
For our calculations, the \qbox~ is calculated including one- and
two-body diagrams at third order in perturbation theory
\cite{Coraggio20c}.

Then, the \qbox~is the building block to solve the non-linear matrix
equation (\ref{eqfinal}) to derive \heff~ through iterative techniques
such as the Kuo-Krenciglowa and Lee-Suzuki ones
\cite{Krenciglowa74,Suzuki80}, or graphical non-iterative methods
\cite{Suzuki11}.

It should be pointed out that the nuclear systems under consideration
for the \dbb~ decay of $^{100}$Mo -- namely the $A=100$ isobars
Mo,Tc,Ru -- are characterized by 22 valence nucleons.
This means that to derive the effective shell-model Hamiltonian and
decay operators, their perturbative expansion should include many-body
diagrams with up to 22 external lines, which provide up to 22-body
components of the shell-model operators.

Since the SM code we have employed for our calculations \cite{KSHELL}
cannot diagonalize \heffs~ that include three-body components, we
adopt a particular procedure when deriving \heff.

We include also diagrams with three external lines in the calculation
of the \qbox, which are responsible of the so-called ``blocking effect'',
that accounts for the Pauli principle within the interaction via the
two-body force of the valence nucleons with configurations outside the
model space.
Then, we have performed a normal-ordering decomposition of the
three-body contributions arising at second order in perturbation
theory, retaining only the two-body term which is then
density-dependent from the number of valence nucleons.
Details of such a procedure can be found in
Refs. \cite{Coraggio20c,Coraggio20e}, together with a study of its
contribution to the eigenvalues of the SM Hamiltonian.

It is worth pointing out that, within a perturbative expansion, that the
relevance of the many-body components of \heff~ is reduced with the
increase of the number of the external valence lines.
As a matter of fact, the lowest-order three-body contribution to
\heff~ shows up to second order of its perturbative expansion, while
the lowest-order of four-body components corresponds to the
third-order one, and so on.
This supports the choice to retain up to the three-body contributions
of many-body induced forces in the derivation of \heff.

We now need to focus on another crucial aspect of our shell-model
calculations, as we have performed them in Ref. \cite{Coraggio22a}.

The model space constituted by 4 and 5 proton and neutron orbitals,
respectively, that we dub as $[45]$, is quite large, and our shell
model code does not allow to construct a sufficiently large number of
intermediate $J^{\pi}=1^+$ states to calculate \nmed~ (see
Eq. \ref{doublebetameGT}).
Then, we consider a smaller model space, where the neutron $0h_{11/2}$
orbital is excluded on the basis of a study of the evolution of the
neutron effective single-particle energies of the Mo isotopes as a
function on the number of valence neutrons.
Such a study, as reported in Ref. \cite{Coraggio22a}, evidences that there
is an energy gap separating the $1d_{5/2},1d_{3/2},2s_{1/2}$ neutron
orbitals from the $0g_{7/2},0h_{11/2}$ ones, which increases as a
function of the number of valence neutrons.

This is the ground to set a new model space, that is dubbed $[44]$,
derived from $[45]$ by cutting out the neutron orbital $0h_{11/2}$.
The following step is to derive a new effective Hamiltonian $H_{\rm
  eff}^{[44]}$, defined in the smaller model space $[44]$, by way of a
unitary transformation of the effective Hamiltonina $H_{\rm
  eff}^{[45]}$, that has been derived in the full model space $[45]$.

This double-step procedure to derive \heff~ for smaller model spaces
has been introduced in Ref. \cite{Coraggio15a}, and an extended
discussion can be found in Ref. \cite{Coraggio16a}, where it has been
shown the effectiveness of the method with respect to a raw truncation
of the original \heff.

\subsection{The SM effective decay operators}\label{oeff}
Here, we discuss the derivation of the SM effective decay operators,
and first we outline the structure of the \dbb-decay one.

\dbb~ decay occurs through two virtual single-$\beta$ transitions, in
terms of the Gamow-Teller (GT) and Fermi decay operators.
The \dbb~ matrix element is then obtained through the calculation of
GT and Fermi components of their nuclear matrix elements as follows:
\begin{eqnarray}
M_{\rm GT}^{2\nu} & = & \sum_n \frac{ \langle 0^+_f || \vec{\sigma} \tau^-
  || 1^+_n \rangle \langle 1^+_n || \vec{\sigma}
\tau^- || 0^+_i \rangle } {E_n + E_0} ~,\label{doublebetameGT} \\
M_{\rm F}^{2\nu} & = & \sum_n \frac{ \langle 0^+_f || \tau^-
  || 0^+_n \rangle \langle 0^+_n || \tau^- || 0^+_i \rangle } {E_n +
  E_0} ~,\label{doublebetameF}
\end{eqnarray}

\noindent
where $E_n$ is the excitation energy of the $J^{\pi}=0^+_n,1^+_n$
intermediate state -- the index $n$ running over all possible
intermediate states induced by the transition operator --, and
$E_0=\frac{1}{2}Q_{\beta\beta}(0^+) +\Delta M$.
$Q_{\beta\beta}(0^+)$ and $\Delta M$ are the $Q$ value of the transition
and the mass difference of the parent and daughter nuclear states,
respectively.

It should be pointed out that, since the Fermi component has a marginal
role with respect to the GT one \cite{Haxton84,Elliott02}, we neglect
it in our calculations.

In order to calculate the expression in Eq. \ref{doublebetameGT}, we
adopt the Lanczos strength-function method \cite{Caurier05}, that is
an efficient way to calculate \nmed, and including a number of
intermediate states that is sufficient to provide the needed accuracy.

The calculated value of \nmed~ can be then compared with the
experimental counterpart, that is extracted from the measured half
life $T^{2\nu}_{1/2}$:

\begin{equation}
\left[ T^{2\nu}_{1/2} \right]^{-1} = G^{2\nu} \left| M_{\rm GT}^{2\nu}
\right|^2 ~,
\label{2nihalflife}
\end{equation}
\noindent
$G^{2\nu}$ being the \dbb-decay phase-space (or kinematic) factor
\cite{Kotila12,Kotila13}.

As we mentioned in the introduction, the decay operators that we
employ to calculate observables such as the $B(E2)$ strengths or the
nuclear matrix element of the \dbb~ decay \nmed, have to account for
the fact that the diagonalization of \heff~ does not provide the real
wave-functions, but their projections onto the model space $P$.
This means that any decay operator $\Theta$ should to take into
account for the neglected degrees of freedom corresponding to the
$Q=1-P$ space.

The problem of deriving SM effective decay-operators through a
perturbative expansion, starting from realistic nuclear potentials,
has been tackled by many authors since late 1960s
\cite{Mavromatis66,Mavromatis67,Federman69,Ellis77,Towner83,Towner87},
and here we follow the procedure that has been introduced by Suzuki and
Okamoto in Ref. \cite{Suzuki95}.
The interesting aspect of such an approach is that the derivation of
the effective decay operators \thetaeff~ is consistent with the
Kuo-Lee-Ratcliff approach to calculate \heff.
In fact, it is based on perturbative expansion of a vertex function
\tbox, analogously with the derivation of \heff~ in terms of the
\qbox~ (see section \ref{heff}).
The details about our application of the Suzuki-Okamoto procedure to
derive SM effective decay operator are reported in
Ref. \cite{Coraggio20c}, and analyses of the convergence properties of
the perturbative expansion of \thetaeff~ have been carried out in
Refs. \cite{Coraggio18,Coraggio20a,Coraggio20c,Coraggio22a}.

In the present work, we are interested in evaluating the contribution
of the two-body component of the GT operator, which we calculate
including diagrams up to second order in perturbation theory.

\begin{figure}[h]
\centerline{\includegraphics[width=6.5truecm]{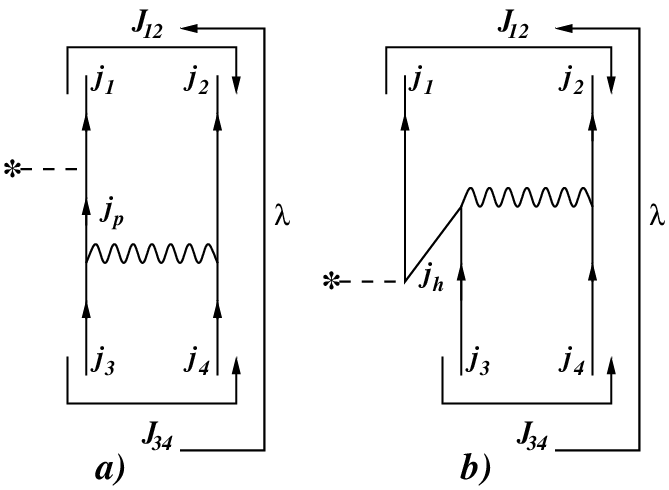}}
\caption{Two-body second-order diagrams included in the perturbative
  expansion of the \tbox. The asterisk indicates the bare operator
  $\Theta$, $\lambda$ is the rank of the decay operator.}
\label{figeffop1}
\end{figure}

In Fig. \ref{figeffop1} we report the diagrams up to second order
appearing in the perturbative expansion of the \tbox~ for a one-body
operator $\Theta$, but for a two valence-nucleon system.
It is important pointing out that for both diagrams {\it a)} and {\it
  b)} occur three other diagrams with the same topology, but the
insertion of the decay-operator $\Theta$ is placed on a different
external valence-particle line.

The calculation of the Goldstone diagrams in Fig. \ref{figeffop1} can
be carried out using the angular momentum coupled representation, as
described by Tom Kuo and coworkers in Ref. \cite{Kuo81}, and their
analytical expressions are the following:

\begin{eqnarray}
a) &=& \hat{J_{34}} (-1)^{J_{12} + J_{34} - \lambda} 
\sum_{j_p} \hat{j_p}  ~ X \left( \begin{array}{ccc} j_p~ j_p~  0  \\ 
j_1~ j_2 ~  J_{12} \\ \lambda ~  J_{34}~  J_{12}  \end{array}\right)
X \left( \begin{array}{ccc} \lambda ~ J_{34} ~  J_{12}  \\ 
 0 ~ J_{34} ~  J_{34} \\ \lambda ~~~  0 ~~~  \lambda  \end{array}\right) \\
~ & \times &\frac{
\langle j_p,j_2; J_{34} | V_{NN} | j_3,j_4; J_{34} \rangle
\langle j_1 || \Theta_{\lambda} || j_p \rangle
}
{
[\epsilon_{0}-(\epsilon_{j_p}+\epsilon_{j_2})]
}~,\nonumber
\label{matela}
\end{eqnarray}

\begin{eqnarray}
b) &=& \hat{J_{34}} (-1)^{J_{12} + J_{34} - \lambda+1} 
\sum_{j_h} \hat{j_h}  ~ X \left( \begin{array}{ccc} j_h~ j_h~  0  \\ 
j_1~ j_2 ~  J_{12} \\ \lambda ~  J_{34}~  J_{12}  \end{array}\right)
X \left( \begin{array}{ccc} \lambda ~ J_{34} ~  J_{12}  \\ 
 0 ~ J_{34} ~  J_{34} \\ \lambda ~~~ 0 ~~~  \lambda  \end{array}\right) \\
~ & \times &\frac{
\langle j_h,j_2; J_{34} | V_{NN} | j_3,j_4; J_{34} \rangle
\langle j_1 || \Theta_{\lambda} || j_h \rangle
}
{
[\epsilon_{0}-(\epsilon_{j_1}-\epsilon_{j_h}+\epsilon_{j_3}+\epsilon_{j_4})]
}~,\nonumber
\label{matelb}
\end{eqnarray}

\noindent
where $\hat{x}=(2x+1)^{1/2}$, $X$ is the so-called standard normalized
9-$j$ symbol \cite{Edmonds57}, $\epsilon_m$ denotes the unperturbed
single-particle energy of the orbital $j_m$, $\epsilon_{0}$ is the
so-called starting energy, namely the unperturbed energy of the
incoming particles $\epsilon_0=\epsilon_{j_3}+\epsilon_{j_4}$ \cite{Kuo81}.
The matrix elements appearing on the right-hand side of
Eqs. (8) and (9) are the two-body matrix elements(TBMEs) of the input
potential $V_{NN}$, that are antisymmetrized but not normalized, and
the reduced matrix elements of the one-body decay operator
$\Theta_{\lambda}$ of rank $\lambda$.

The second-order two-body contributions to the effective $\beta$-decay
operators have been considered in two recent papers
\cite{Coraggio24b,Degregorio24}, and in this work we are going to show
their impact with respect to arresting the perturbative expansion of
\thetaeff~ to one-body contributions only.

\section{GT strengths and \nmeds~ of $^{100}$Mo decay}\label{results}
In our previous study of $^{100}$Mo \dbb~ decay \cite{Coraggio22a},
we have already shown the ability of our \heff~ and effective
electromagnetic-transition operators to reproduce the low-energy
spectroscopic properties of $^{100}$Mo and $^{100}$Ru, the latter
being the granddaughter nucleus involved in the decay under present
investigation.

Then, we start presenting the calculated quantities which are directly
linked to the \dbb~ decay of $^{100}$Mo.
It should first be noticed that the selection rules of the GT operator
connect only the proton and neutron orbitals $0g_{9/2},0g_{7/2}$.
Consequently, the one-body effective GT$^+$ operator consists of a single
matrix element that corresponds to the decay $\pi 0g_{9/2} \rightarrow
\nu 0g_{7/2}$, and it corresponds to a calculated quenching factor $q=0.454$.
Correspondingly, the sole matrix element of the single-body component
of the effective GT$^-$ operator $\nu 0g_{9/2} \rightarrow \pi
0g_{7/2}$ tallies with a quenching factor $q=0.503$.

Note that the non-hermiticity of the effective GT-decay operator
traces back to two different sources.
First, the proton and neutron model spaces we have chosen are
different, and the proton-neutron symmetry is broken because of
different sets of particle-hole intermediate configurations appear in
the perturbative expansion of the effective GT operator where
moreover, we account also for the Coulomb interaction between
protons.
Second, the Suzuki-Okamoto procedure that we follow to derive the
effective operators is non-hermitian \cite{Suzuki95}.

We start by comparing the experimental running sums of the $^{100}$Mo
$\sum {\rm B(GT)}$ strengths with the ones calculated by employing the
bare GT$^-$ operator and the effective GT$^-$ operators, with and
without two-body components.

It is worth reminding that the experimental GT strengths obtained from
charge-exchange reactions are extracted from the GT component of the
cross section at zero degree, following the standard approach in the
distorted-wave Born approximation (DWBA):

\[
\frac{d\sigma^{GT}(0^\circ)}{d\Omega} = \left (\frac{\mu}{\pi \hbar^2} \right
)^2 \frac{k_f}{k_i} N^{\sigma \tau}_{D}| J_{\sigma \tau} |^2 B({\rm GT})~~,
\]

\noindent
where $N^{\sigma \tau}_{D}$ is the distortion factor, $| J_{\sigma
  \tau} |$ is the volume integral of the effective $NN$ interaction,
$k_i$ and $k_f$ are the initial and final momenta, respectively, and
$\mu$ is the reduced mass (see formula and description in
Refs. \cite{Puppe12,Frekers13}).
This means that those that are indicated as ``the experimental GT
strengths'' might be considered somehow model-dependent.

\begin{figure}[h]
\centerline{\includegraphics[width=7.0truecm,angle=-90]{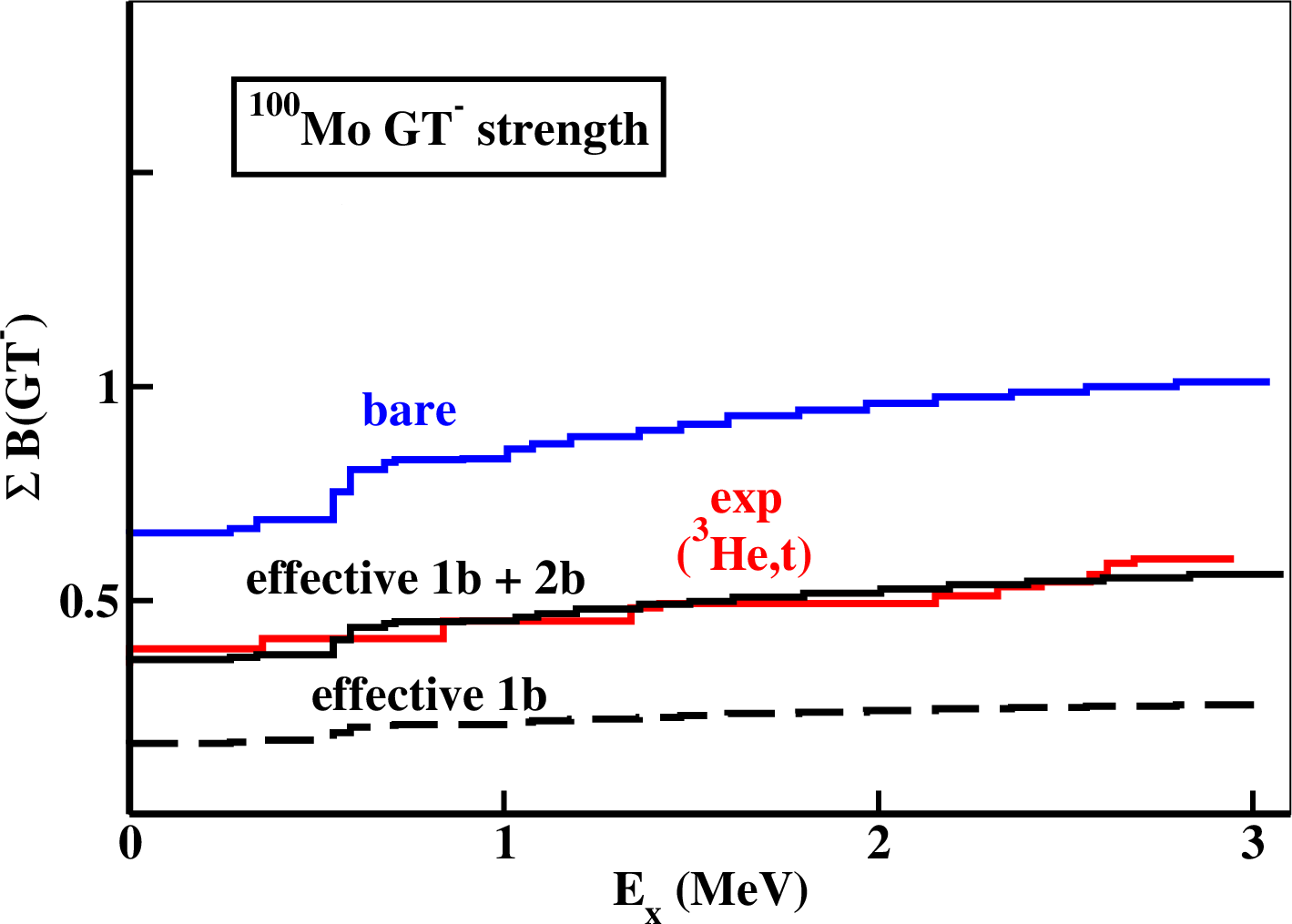}}
\caption{Running sums of the $^{100}$Mo $\sum {\rm B(GT^-)}$
  strengths as a function of the excitation energy $E_x$ up to 3 MeV.}
\label{100MoGT-}
\end{figure}

In Fig. \ref{100MoGT-} we have reported the calculated $\sum {\rm
  B(GT^-)}$ for $^{100}$Mo as a function of the $^{100}$Tc excitation
energy up to 3 MeV, and compared with the available data reported with
a red line \cite{Thies12c}.
The results obtained with the bare operator are drawn with a blue
line, those obtained employing only the one-body component of the
effective GT$^-$ operator are plotted with a  dashed black line, and
those calculated with both the one- and two-body components of the
effective decay operator are reported with a continuous black line.

From the inspection of  Fig. \ref{100MoGT-}, we observe that the
distribution obtained using the bare operator overestimates the
observed one, while the quenching induced by the single-body effective
operator leads to an underestimation of the values extracted
from the experiment.
A better reproduction of the observed running sums is then obtained
considering both one- and two-body components of the GT$^-$ effective
operator: this is an expected feature of the two-body component of the
effective operator, since it should reduce, when increasing the number of
the valence nucleons, the renormalization of the bare operator,
leading to the so-called ``blocking effect'' \cite{Ellis77}.

Then, to provide a quantitative evaluation of the role of the
different contributions to the effective decay operator, we point out
that the theoretical total GT$^-$ strengths are 1.082, 0.274, and
0.611 with the bare operator, the effective one-body operator, and the
effective operator including both one- and two-body components,
respectively.
This quantifies the non-negligible role of the ``blocking effect'' in a
nuclear system, such as $^{100}$Mo, which is characterized by 14 and 8
valence protons and neutrons, respectively.

Now, we shift the focus on the calculation of the nuclear matrix
elements of the \dbb~ decay of $^{100}$Mo.

\begin{table}[h]
\tbl{Experimental \cite{Barabash20}  and calculated \nmeds~(in
    MeV$^{-1}$) for $^{100}$Mo \dbb~decay. The theoretical values are
    obtained employing both the bare (I), effective one-body  (II),
    and effective one- and two-body \dbb~ operators (III).}
{\begin{tabular}{@{}ccccc@{}} \toprule
   \label{ME_100Mo}
   $^{100}$Mo$\rightarrow^{100}$Ru decay branches & Experiment & I & II & III \\
\colrule
$J^{\pi}=0^+_1\rightarrow J^{\pi}=0^+_1$ & $0.224 \pm 0.002$ & 0.896 & 0.205 & 0.374 \\
$J^{\pi}=0^+_1\rightarrow J^{\pi}=0^+_2$ & $0.182 \pm 0.006$ & 0.479 & 0.109 & 0.135 \\
\botrule
\end{tabular}}

\end{table}

The experimental and calculated values of the \nmeds are reported in
Table 1 for the decay from the $J^{\pi}=0^+_1$ $^{100}$Mo
ground state (g.s.) to the $^{100}$Ru $J^{\pi}=0^+_1,0^+_2$ states.

We note that, for both decays, the values of \nmed~ obtained with the
bare operator (indicated with the label I) overestimate the
experimental ones, and employing the matrix elements of the effective
single-body component of the GT$^+$ and GT$^-$ operators (II) we
obtain results that are in a far better agreement with the observed \nmeds.
As for the case of the GT$^-$ running sums, the impact of the two-body
component of the effective GT-decay operator (III) is sizeable because
of the large number of valence nucleons, and provides the expected
effect of softening the quenching factors obtained from the one-body
component of the effective operator.

\section{Conclusions}
This contribution is meant to celebrate the research work of Tom Kuo,
and his achievements that have been of paramount relevance for those
researchers who investigate about the problem of handling realistic
nuclear forces in nuclear structure calculations, and of deriving
effective Hamiltonians and decay operators in restricted model
spaces.

To this end, we have discussed the role of the many-body components of
shell-model effective decay-operators, which are characterized by a
one-body nature as a bare operator (e.g. electromagnetic multipole
transitions, GT  decay, etc.).
These terms, that contribute to the construction of a many-body decay
operator, may be relevant for nuclear systems with a number of valence
nucleons larger than one, namely they account for the reduction of
accessible configurations because of the Pauli principle for identical
nucleons.

We have started from the one-body GT-decay operator, and constructed,
within the many-body perturbation theory, a SM effective decay-operator
owning both one- and two-body components.
Then, we have evaluated the impact of the two-body component of the
effective operator with respect to the one-body one, considering the
double-$\beta$ decay of $^{100}$Mo, a process which in the model space
above $^{78}$Ni core involves 22 valence nucleons.

As shown in Sec. \ref{results}, the two-body component of the
effective operator reduces the quenching generated by the one-body
component by $50\%$ for the total GT$^-$ strength, and up to $35\%$
for the quenching of the \dbb~ nuclear matrix elements.
As expected, this is a non-negligible effect in a decay of a nucleus
with many valence particles, and indicates that the theoretical
description of the renormalization of the GT-decay operator does not
support the existence of an ``universal'' quenching factor $q$, since
this depends largely on the model space and the number of interacting
nucleons.

We want to conclude by witnessing that the profound impact of Tom Kuo
on the field of theoretical nuclear structure, and on those who had
the privilege to work and learn from him cannot be overstated.

His pioneering research, unwavering dedication, and generous
mentorship have left an indelible mark on the scientific community and
have significantly advanced our understanding of nuclear structure in
terms of realistic nuclear potentials.

It is with deep respect and gratitude that we honor Tom Kuo's legacy,
and dedicate this work to his memory, acknowledging that his spirit of
inquiry and passion for discovery will inspire and guide future
generations of nuclear physicists.

\section*{Acknowledgements}
The authors acknowledge the financial support of the European Union -
Next Generation EU, Mission 4 Component 1, CUP I53D23001060006, for
the PRIN 2022 project "Exploiting separation of scales in nuclear
structure and dynamics".

\bibliographystyle{ws-ijmpe}
\bibliography{biblioWS.bib}



\end{document}